\documentclass[11pt]{article}
\usepackage{graphicx}
\usepackage{amsmath,amssymb}
\def\1{{\bf 1}}

\def\be{\begin{equation}}
\def\ee{\end{equation}}

\begin{document}
\noindent {\large \bf Optimal atomic-resolution structures of prion AGAAAAGA amyloid fibrils}\\

\bigskip

\noindent {\Large Jiapu Zhang$^{1*}$, Jie Sun$^2$, Changzhi Wu$^3$}\\

\noindent $^{1*}$School of Sciences, Information Technology, and Engineering,\\
The University of Ballarat, Mount Helen Campus, Victoria 3353, Australia,\\
Phone: {\it 61-423487360}, Email: {\it j.zhang@ballarat.edu.au, jiapu\_zhang@hotmail.com}\\
\noindent $^2$Provost's Chair Professor, National University of Singapore,
Singapore 119245\\
\noindent $^3$Optimal Control Professor, Chongqing Normal University,
Chongqing 630047, China\\

\noindent $^*$to whom correspondence should be addressed\\

\vskip 0.6cm

\noindent {\bf \Large Abstract}\\
{\small 
X-ray crystallography is a powerful tool to determine the protein 3D structure. However, it is time-consuming and expensive, and not all proteins can be successfully crystallized, particularly for membrane proteins. Although nuclear magnetic resonance (NMR) spectroscopy is indeed a very powerful tool in determining the 3D structures of membrane proteins, it is also time-consuming and costly. To the best of the authors' knowledge, there is little structural data available on the AGAAAAGA palindrome in the hydrophobic region (113–-120) of prion proteins due to the noncrystalline and insoluble nature of the amyloid fibril, although many experimental studies have shown that this region has amyloid fibril forming properties and plays an important role in prion diseases. In view of this, the present study is devoted to address this problem from computational approaches such as global energy optimization, simulated annealing, and structural bioinformatics. The optimal atomic-resolution structures of prion AGAAAAGA amyloid fibils reported in this paper have a value to the scientific community in its drive to find treatments for prion diseases.\\
}


\noindent ---------------------------------------------------------------------------------------------------------------\\
This paper was first reported in the Proceedings of the International Conference on Optimization and Control 2010, July 2010, Perth/Chongqing/Guiyang, pp. 91--112.
This paper is dedicated to Professor Kok Lay Teo (Curtin University of Technology, Australia) and Professor Jie Sun (National University of Singapore, Singapore) on the occasion of their 65th birthday, to Professor Changyu Wang (Qufu Normal University, China) on the occasion of his 75th birthday, and to Professor Masao Fukushima on the occasion of his 60th birthday.

\newpage
\section{Introduction}
Prion diseases include Creutzfeldt-Jakob disease, variant Creutzfeldt-Jakob diseases, Gerstmann-Strussler-Scheinker syndrome, Fatal Familial Insomnia, Kuru in humans, scrapie in sheep, bovine spongiform encephalopathy (or 'mad-cow' disease) and chronic wasting disease in cattle. These diseases are invariably fatal and highly infectious neurodegenerative diseases affecting humans and animals. However, for threating all these diseases, there is no effective therapeutic approaches or medications [Aguzzi and Heikenwalder, 2006, Prusiner, 1998, Weissmann, 2004]. %\cite{aguzzi2006, prusiner1998, weissmann2004}. 
Prion diseases are caused by conversion of prion proteins from a soluble into an insoluble fibrillar form \cite{prusiner1982, prusiner1998}. The normal cellular prion protein (PrP$^{\text{C}}$) is rich in $\alpha$-helices but the infectious prions (PrP$^{\text{Sc}}$) are rich in $\beta$-sheets \cite{griffith1967}. For PrP$^{\text{C}}$, the N-terminal residues (1--123) are unstructured, but the C-terminal residues (124--231) are well structured with three $\alpha$-helices, two short anti-parallel $\beta$-strands, and a buried disulfide bond between the 2nd and 3rd $\alpha$-helix (residues number 179 and number 214). The prion protein molecular structure and dynamics can be seen from \cite{zhang2010, zhang2011b, zhang2011c} etc.. For PrP$^{\text{Sc}}$, the rich $\beta$-sheets formulate into prion amyloid fibrils.\\ 

The hydrophobic region (113–-120) AGAAAAGA palindrome of prion proteins falls just within the N-terminal unstructured region (1--123) of prion proteins which is hard to determine its molecular structure using NMR spectroscopy or X-ray crystallography \cite{riek1996}. However, many experimental studies such as 
[Brown, 2000, Brown, 2001, Brown et al., 1994, Holscher et al., 1998, Jobling et al., 2001, Jobling et al., 1999, Kuwata et al., 2003, Norstrom and Mastrianni, 2005, Wegner et al., 2002]
%\cite{brown2000,brown2001,brown1994,holscher1998,jobling2001,jobling1999,kuwata2003,norstrom2005,wegner2002} 
have shown that: (1) the hydrophobic region (113–-120) AGAAAAGA of prion proteins plays an important role in the conversion of PrP$^{\text{C}}$ to the abnormally folded form PrP$^{\text{Sc}}$; and (2) AGAAAAGA is important for amyloid fibril formation and is an inhibitor of prion diseases. Zhang (2009) also confirmed through computer molecular dynamics simulations that the stability of prion proteins might be attributable mainly to the N-terminal unstructured region (1--123). Due to the noncrystalline and insoluble nature of the amyloid fibril, it is very difficult to obtain atomic-resolution structures of AGAAAAGA using traditional experimental methods \cite{tsai2005,zheng2006}. For the sake of clarity on computers again, we use the program used by \cite{zhang2007} to theoretically confirm that prion (113–-120) AGAAAAGA segment has an amyloid fibril forming property. The theoretical computation results are shown in Fig. 4 of \cite{zhang2011a}, from which we can see that the prion AGAAAAGA region (113–-120) is clearly identified as the amyloid fibril formation region because the energy is less than the amyloid fibril formation threshold energy of -26 KCal/mol \cite{zhang2007}. Thus, we got confidence in constructing the atomic-resolution molecular structures of prion (113--120) AGAAAAGA amyloid fibrils by computer computational approaches or introducing novel mathematical formulations and physical concepts.\\

Many studies have indicated that computational approaches or introducing novel mathematical formulations and physical concepts into molecular biology, such as Mahalanobis distance \cite{chou1995, chou_zhang1995}, pseudo amino acid composition \cite{chou2001, chou2011}, graphic rules \cite{andraos2008, chou1989b, chou1990, chou2010, zhou1984}, complexity measure factor [Xiao et al., 2006b, Xiao et al., 2005b], 
%\cite{xiao2006b,xiao2005b}, 
homology modeling \cite{chou2004a}, cellular automaton [Xiao and Chou, 2007, Xiao et al., 2006a, Xiao et al., 2009, Xiao et al., 2005a],
%\cite{xiao_chou2007, xiao2006a,xiao2009,xiao2005a}, 
molecular docking \cite{chouetal2003}, grey theory \cite{xiaoetal2008a}, geometric moments [Xiao et al., 2008b],
%\cite{xiao2008b}, 
low-frequency (or Terahertz frequency) phonons \cite{chou1988,chou1989a,chou_chen1977}, solitary wave \cite{chouetal1994,sinkala2006}, and surface diffusion-controlled reaction \cite{chou_zhou1982}, can significantly stimulate the development of biological and medical science. Various computer computational approaches were used to address the problems related to ``amyloid fibril" [Carter and Chou, 1998, Chou, 2004b, Chou, 2004c, Chou and Howe, 2002, Wang et al., 2008, Wei et al., 2005].
%\cite{carter1998, chou2004b, chou2004c, chou2002, wang2008, wei2005}. 
Here, we would like to use the hybrid local and global optimization search methods to investigate the optimal atomic-resolution amyloid fibril models in hopes that the findings thus obtained may be of use for controlling prion diseases. Using the traditional local search steepest descent and conjugate gradient [Li and Chen, 2005, Sun and Zhang, 2001, Zhu and Chen, 2008]
%\cite{li2005,sun2001,zhu2008} 
methods hybridized with the standard global search simulated annealing method [Horst et al., 2003, Yiu et al., 2004],
%\cite{horst2003,yiu2004}, 
zhang (2011a) successfully constructed three optimal atomic-resolution structures of prion AGAAAAGA amyloid fibrils. These structures were constructed based on the breakthrough work of \cite{sawaya2007}. In \cite{zhang2011a}, the author pointed out that basing on the NNQNTF peptide of elk prion 173-178 (PDB entry 3FVA that was released into Protein Data Bank (http://www.rcsb.org) on 30-JUN-2009, deposition date  15-JAN-2009) we might also be able to construct amyloid fibril models for prion AGAAAAGA palindrome; this paper is doing this homology model construction work. The homology models were built using an improved hybrid Simulated Annealing (SA \cite{kirkpatrick1983}) Discrete Gradient (DG \cite{bagirov2008, bagirov2003}) method. Then the models were optimized/solved using the traditional steepest descent (SD) and conjugate gradient (CG) local search methods of \cite{case2008}; the former has nice convergence but is slow when close to minimums and the latter is efficient but its gradient RMS and GMAX gradient do not have a good convergence \cite{case2008}. We used the SD method followed by the CG method to optimize our models. When the models could not be optimized further, we employed the standard global search SA method of \cite{case2008} to escape from the stationary point calculated by the local search SD \& CG methods. Through the further refinement of SD and CG local search methods, at last two optimal models were successfully got. Numerical results in this paper show that the hybridization of local and global search optimization methods is very effective. X-ray crystallography finds the X-ray final structure of a protein, which usually need refinements using a SA protocol in order to produce a better structure; this paper also correctly illustrates the SA protocol of crystallography.\\

X-ray crystallography is a powerful tool to determine the protein 3D structure. However, it is time-consuming and expensive, and not all proteins can be successfully crystallized, particularly for membrane proteins. Although NMR spectroscopy is indeed a very powerful tool in determining the 3D structures of membrane proteins [Call, et al., 2010, Call, et al., 2006, Oxenoid and Chou, 2005, Pielak and Chou, 2010a, Pielak and Chou, 2010b, Pielak and Chou, 2011, Pielak et al., 2009, Schnell and Chou, 2008, Wang et al., 2009],
%\cite{call2010, call2006,oxenoid2005,pielak2010a, pielak2010b, pielak2011, pielak2009, schnell2008, wang2009}, 
it is also time-consuming and costly. To the best of the authors' knowledge, there is little structural data available on the AGAAAAGA palindrome in the hydrophobic region (113--120) of prion proteins, although many experimental studies have shown that this region has amyloid fibril forming properties. In view of this, the present study was devoted to address this problem from computational approaches such as global energy optimization \cite{chou1982,chou1992b}, simulated annealing \cite{chou1992a,chou1991}, and structural bioinformatics \cite{chou2004c}. Thus, the optimization computational approaches (such as SADG, GADG, SDCG etc) presented in this paper are very necessary and important to study amyloid fibrils, nanotubes, etc.. The prion AGAAAAGA optimal atomic-resolution structures of this paper might have a value for finding treatments for prion diseases.

\section{The Optimization Model Building}
Recently the protein fibril structure of NNQNTF (173-178) segment from elk prion protein was released \cite{wiltzius2009}. Its PDB entry ID is 3FVA in the Protein Data Bank. This fibril has six chains, belonging to Class 1 of \cite{sawaya2007}. The atomic structure is a steric zipper, with strong van der Waals (vdw) interactions between $\beta$-sheets and hydrogen bonds to maintain the $\beta$-strands.\\ 

Basing on this steric zipper, two prion AGAAAAGA palindrome amyloid fibril models –- a six chains AAAAGA model (Model 1) and a six chains GAAAAG model (Model 2) –- will be successfully constructed. The minimum sequence necessary for fibril formation should be AGAAA, AAAGA, AGAAAA, GAAAAG, AAAAGA, AGAAAAG, GAAAAGA or AGAAAAGA \cite{zhang2011a}, which are important for fibril formation and are an inhibitor of PrP$^{\text{Sc}}$ neurotoxicity \cite{brown2000}. Because the peptide NNQNTF has six residues and the six chains AGAAAA model could not successfully pass SA, AAAAGA and GAAAAG were picked out of the eight possible sequences. The D chain (i.e. $\beta$-sheet 2) of 3FVA.pdb can be obtained from A Chain (i.e. $\beta$-sheet 1) using the mathematical formula
\begin{equation}
D = \left( \begin{array}{ccc}
-1  &0  &0  \\
 0  &1  &0     \\
 0  &0  &-1 \end{array} \right) A + 
\left( \begin{array}{c}
-14.31482\\
2.42\\
-21.03096 
\end{array} \right).
\end{equation}
AD chains of Models 1-2 (Figures 1-2) were respectively got from AD chains of 3FVA.pdb using the mutate module of the free package Swiss-PdbViewer (SPDBV Version 4.01) ({\small http://spdbv.vital-it.ch}), but the vdw contacts are too far, very bad at this moment (Figures 1-2). To get good vdw interactions will be an optimization problem described as follows.

\begin{figure}[h!]
\centerline{
\includegraphics[scale=0.45]{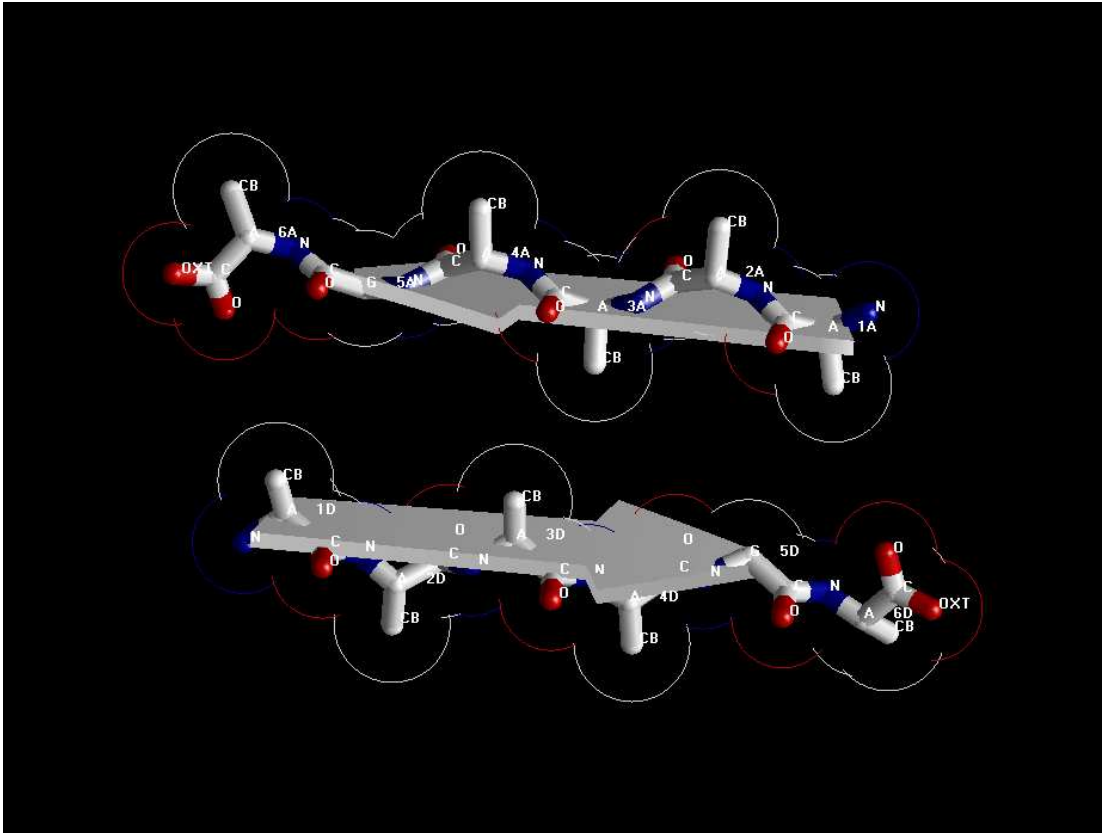}
}
\caption{Model 1 - bad vdw contacts of AD chains of the AAAAGA model.}
\label{fig3}
\end{figure}

\begin{figure}[h!]
\centerline{
\includegraphics[scale=0.45]{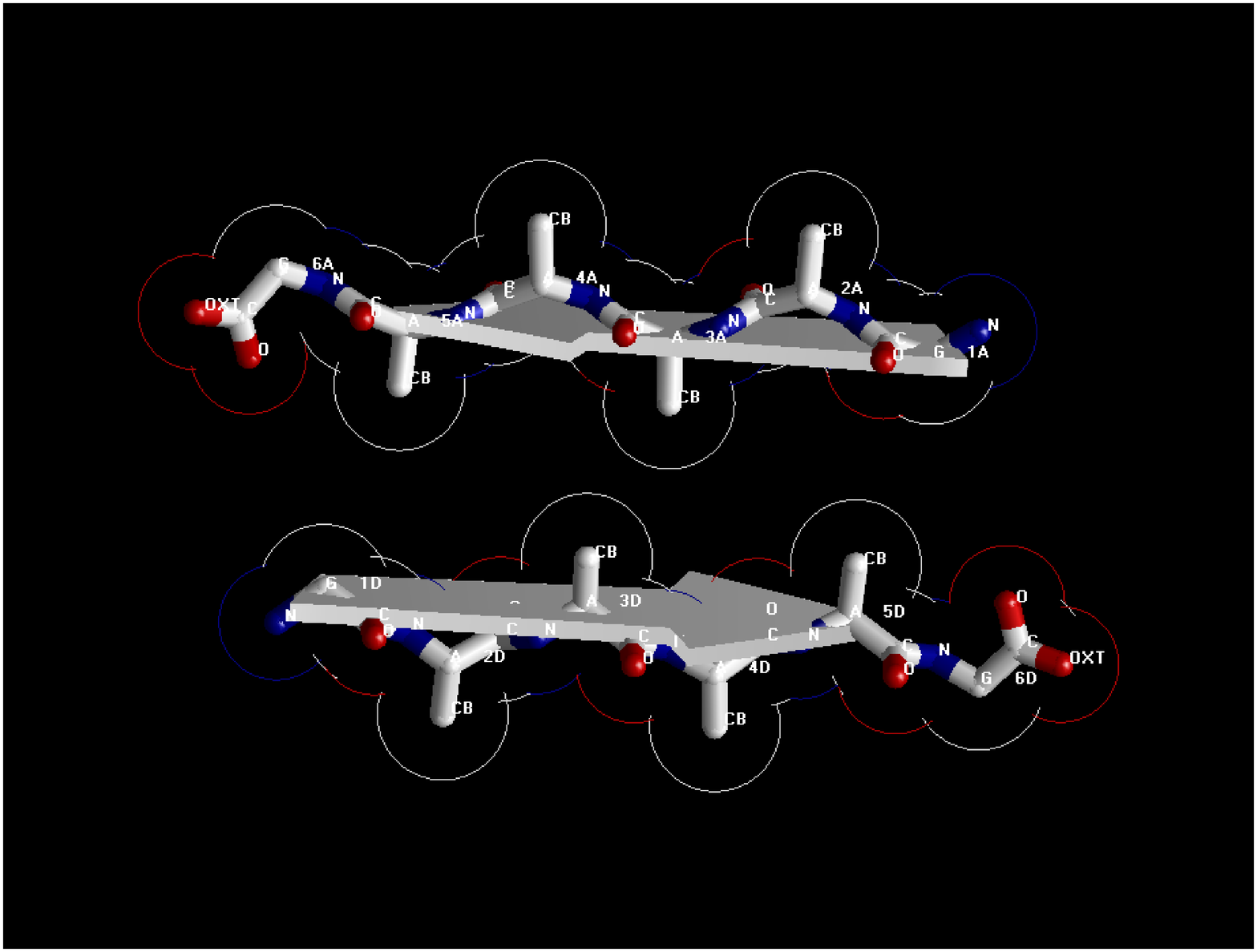}
}
\caption{Model 2 - bad vdw contacts of AD chains of the GAAAAG model.}
\label{fig4}
\end{figure}

Neutral atoms are subject to two distinct forces in the limit of large distance and short distance: a dispersion force (i.e. attractive vdw force) at long ranges, and a repulsion force, the result of overlapping electron orbitals. The Lennard-Jones (L-J) potential represents this behavior ({\small http://en.wikipedia.org/wiki/Lennard-Jones\_potential}, or Locatelli M. et al. (2008) and references therein). The L-J potential is of the form
\begin{equation} \label{LJ_r_form}
V(r)=4\varepsilon \left[ (\frac{\sigma}{r})^{12} - (\frac{\sigma}{r})^6 \right],
\end{equation}
where $\varepsilon$ is the depth of the potential well and $\sigma$ is the atom diameter; these parameters can be fitted to reproduce experimental data or deduced from results of accurate quantum chemistry calculations. The $(\frac{\sigma}{r})^{12}$ term describes repulsion and the $(\frac{\sigma}{r})^6$ term describes attraction. If we introduce the coordinates of the atoms whose number is denoted by $N$ and let $\varepsilon = \sigma =1$ be the reduced units, the form (\ref{LJ_r_form}) becomes
\begin{equation}\label{LJ_x_form}
f(x)=4\sum_{i=1}^N \sum_{j=1,j<i}^N \left( \frac{1}{\tau_{ij}^6}
-\frac{1}{\tau_{ij}^3} \right),
\end{equation}
\noindent where $\tau_{ij}=(x_{3i-2}-x_{3j-2})^2
                          +(x_{3i-1}-x_{3j-1})^2
                          +(x_{3i}  -x_{3j}  )^2$,
$(x_{3i-2},x_{3i-1},x_{3i})$ is the coordinates of atom $i$, $N \geq 2$. The minimization of L-J potential $f(x)$ on $\mathbb{R}^n$ (where $n=3N$) is an optimization problem:
\begin{equation}\label{LJ_f_form}
\min_{s.t. x\in \mathbb{R}^{3N}} f(x).
\end{equation}

For solving the optimization problem (4), many studies, for examples [Coleman et al., 1994, Doye, 1999, Huang et al., 2002, Huang and pardalos, 2002, Leary, 1997, Wolfe, 1975, Wolfe, 1976, Pardalos et al., 1994, Romero et al., 1999, Xue et al., 1992, Xue, 1993, Xue, 1994a, Xue, 1994b, Xue, 2002],
%\cite{coleman1994,doye1999,huang2002a,huang2002b,leary1997,wolfe1975,wolfe1976,pardalos1994,romero1999,xue1992, xue1993, xue1994a, xue1994b, xue2002},
have been done. A brief review on some results and computational algorithms before the year 2003 for solving (4) can be referred to \cite{zhang2003}, for the year 2004 can be referred to \cite{xiang2004a, xiang2004b}; the website http://www-wales.ch.cam.ac.uk/index.html should not be ignored. Basing on these studies, we present the following two successful algorithms.\\

\noindent {\sf Algorithm 1: Hybrid SADG method}\\
{\small {\sf {\bf \underline{Initialization:}}
\begin{enumerate}
\item[] {\it Define the objective function $f$ and its feasible solution space}.
\item[] Call the {\it initial feasible solution generating procedure} to get $x$.
\item[] Call {\it initial temperature selecting procedure} to get $T$.
\item[] Inialization of $f$: $f=f(x)$.
\item[] Initialize the neigbourhood feasible solution: $x\_{neighbour}=0$.
\item[] Initialize $x\_{best}$: $x\_{best}=x$.
\item[] Initialize $f\_{best}$: $f\_{best}=f$.
\end{enumerate}
do $\{$
\begin{enumerate}
\item[] {\bf \underline{DG local search part:}}
\item[] $\qquad f\_{best}\_{local} = {\bf local\_{search}} (x\_{best},
x\_{new}\_{gotten})$;
\item[] $\qquad x=x\_{new}\_{gotten}$;
\item[] {\bf \underline{SA global search part:}}
\begin{enumerate}
\item[] do $\{$
\item[] $\qquad$ do $\{$
           \begin{enumerate}
           \item[] $\qquad$ $x\_{neighbour}=randomly\_{perturb}(x)$ \cite{bagirov_jiapu2003};
           \item[] $\qquad$ $f\_{neighbour}=f(x\_{neighbour})$;
           \item[] $\qquad $ Calculate the difference $\Delta=f\_{neighbour}-f$;
           \item[] $\qquad$ {\bf If} ($\Delta \leq 0$) {\bf or}
                   (random[0,1] $<$ exp(-$\Delta /T$))
           \item[] $\qquad \qquad x=x\_{neighbour} \quad f=f\_{neighbour}$;
           \item[] $\qquad$ {\bf If} ($f \leq f\_{best})$ $\quad$
                   $\quad x\_{best}=x \quad f\_{best}=f$;
           \end{enumerate}
\item[] $\qquad$ $\}$ while (equilibrium has not been reached);
\item[] $\qquad$ Temperature annealing
\item[] $\}$ while (SA stop criterion has not been met);
\end{enumerate}
\end{enumerate}
$\}$ while ( $f\_{best} - f\_{best}\_{local} \leq -0.001$ ); } }

\noindent The local search DG method is efficient and effective but it is also always trapped in a local solution. SA is a global search method but sometimes just gets a global solution with low probability according to the Metropolis criterion. SA can make DG jump out of the local trap and then DG can make SA reach an optimal solution with the 100\% full probability. The cons of Algorithm 1 might be the large numbers of iterations in SA when searching the global solution with low probability.\\

\noindent {\sf Algorithm 2: Hybrid GADG (Genetic Algorithm \cite{forrest1993} DG) method}
{\small {\sf 
\begin{enumerate}
\item[] {\it Step 0.} Set the seeding of the initial parental population. We set the cluster of 98 atoms as the base of the seed because the 98 atoms cluster has a tetrahedral structure.
\item[] {\it Step 1.} Apply the discrete gradient method on all individuals of the initial parental population to relax them to their nearest local minimal energy positions.
\item[] {\it Step 2.} Call the mating procedure of \cite{deaven1995} to get the center of the mass of the parental population. Then set a random number for this mating procedure to mate more parents with each other. Thus the offspring population is produced.
\item[] {\it Step 3.} Select from the parental population and offspring population to get the best combination of a new population. Take the new population as parental population.
\item[] {\it Step 4.} Run the Newton method (where the Hessian matrix is calculated explicitly) [Dennis et al., 1996]
%\cite{dennis1996} 
to relax all the individuals of parental population to local minimal energy positions.
\item[] {\it Step 5.} Run discrete gradient method to refine the local minimum positions. The refined local minimal energy positions are set as the parental population.
\item[] {\it Step 6.} Call the twinning mutations of \cite{wolf1998}. Then set a random number as the mutation rate for these mutation schemes to make mutations to the parental population. Then offspring population is produced.
\item[] {\it Step 7.} Make a best combination of the parental population and the offspring population to get a new population. Take the new population as parental population.
\item[] {\it Step 8.} Run the explicit Newton method to relax all the individuals of parental population to local minimal energy positions.
\item[] {\it Step 9.} Repeat Step 5.
\item[] {\it Step 10.} Repeat Step 7.
\item[] {\it Step 11.} If the algorithm reaches its convergence, then terminates, otherwise, goto Step 2.
\end{enumerate}
}
}
\noindent In Algorithm 2 the DG method and the explicit Newton method are local search optimization methods, and genetic algorithm is a global search method that brings the local solution to jump out of the trap. The cons of the Algorithm 2 lie in its initial solution specially chosen; whereas Algorithm 1 can start from any initial solution. The computational load of Algorithm 2 is very heavy, compared with that of Algorithm 1.     

SA and GA methods both are stochastic heuristic global search methods. But there are three basic differences between SA and GA: (1) SA simulates the annealing process of crystal materials with Monte Carlo property, GA simulates the process of natural competitive selection, crossover, and mutation of species; (2) GA is a parallel computing (population) algorithm and SA algorithm is a sequential computing (individual) algorithm; (3) The choice of initial solution is different. In Step 0 of Algorithm 2, we may choose clusters of 55, 75, 76, 85, 97, 98, or 147 atoms as the base of seed, because of their well-known structures of the best known solutions. For example, the 98 atoms cluster has a tetrahedral structure; the 55 atoms cluster has an icosahedral structure; the cluster of 75 or 85 atoms has an octahedral structure; and the cluster of 76 atoms has a decahedral structure. For 48 atoms, we use the structures of 13, 23, 19, 10 atoms. In brief, all those 38 atoms face-centred-cubic (fcc) truncated octahedron, 75--77 and 102--104 atoms decahedron, and 98 atoms tetrahedron structures are based on polytetrahedral packing, which is structurally more similar to the icosahedral packing of most Lennard-Jones clusters. Because of different background of SA and GA, the acceptance probability of solution is different very much. The computation load of GADG is clearly large than that of SADG.     

In Algorithm 1 the DG method \cite{bagirov2008} is a derivative-free local search method for nonsmooth optimization with the continuous approximations to the Clarke subdifferential \cite{clarke1983} and the Demyanov-Rubinov quasidifferential [Demyanov and Rubinov, 2000],
%\cite{demyanov2000}, 
and the SA algorithm is using the neighborhood search procedure of \cite{bagirov_jiapu2003}. The convergence of the proposed hybrid method directly follows from the convergence of SA and DG methods. The hybrid method starts from an initial point, first executes DG method to find local minimum, then carries on SA method in order to escape from this local minimum and to find a new starting point for the DG method. Then we again apply the DG method starting from the last point and so on until the sequence of the optimal objective function values gotten is convergent. Similarly for Algorithm 2, the DG and Newton methods are local search optimization methods, GA is global search optimization method, and the convergence of the hybrid GADG method simply follows from that of these three optimization methods. Algorithm 2 can successfully reproduce all the best L-J potential energy values known ({\small http://physchem.ox.ac.uk/$\sim$doye/jon/structures/LJ.html}) nearly up to 310 atoms and fortunately some more precise energy values and better solution structures are got (Table 1 and Figures 3-4), but it is not easy to be applied to our model construction work of this paper.

%\noindent {\bf Table 1} Our precise L-J potential energy best values:
\begin{table}[h!]
\begin{center}
\caption{Our precise L-J potential energy best values}
\label{table1}\small{
\begin{tabular}{cll}
\hline Number of atoms &Our precise best value &Best value known$^*$\\ \hline 
       39              &-180.033185202447      &-180.033185140508\\ \hline
       40              &-185.249838614238      &-185.249838598471\\ \hline
       42              &-196.277533506901      &-196.277533404920\\ \hline
       48              &-232.199531999140      &-232.199529316227\\ \hline
       55              &-279.248470461822      &-279.248470308143\\ \hline
       75              &-397.492330708363      &-397.492330681104\\ \hline
       76              &-402.894865906469      &-402.894865881675\\ \hline
       97              &-536.681382651509      &-536.681382483245\\ \hline
\end{tabular}}
$^*${\small http://physchem.ox.ac.uk/$\sim$doye/jon/structures/LJ.html}
\end{center}
\end{table}

%\newpage
%\begin{figure*}[h!]
%\centerline{
%\includegraphics[scale=0.26]{Fig03_01_39atoms.eps}
%\includegraphics[scale=0.26]{Fig03_02_40atoms.eps}
%\includegraphics[scale=0.26]{Fig03_03_39atoms_line.eps}
%\includegraphics[scale=0.26]{Fig03_04_40atoms_line.eps}
%}
%\label{fig3_39_40}
%\end{figure*}
%\begin{figure*}[h!]
%\centerline{
%\includegraphics[scale=0.26]{Fig03_05_42atoms.eps}
%\includegraphics[scale=0.26]{Fig03_06_48atoms.eps}
%\includegraphics[scale=0.26]{Fig03_07_42atoms_line.eps}
%\includegraphics[scale=0.26]{Fig03_08_48atoms_line.eps}
%}
%\label{fig3_42_48}
%\end{figure*}
%\begin{figure*}[h!]
%\centerline{
%\includegraphics[scale=0.26]{Fig03_09_55atoms.eps}
%\includegraphics[scale=0.26]{Fig03_10_75atoms.eps}
%\includegraphics[scale=0.26]{Fig03_11_55atoms_line.eps}
%\includegraphics[scale=0.26]{Fig03_12_75atoms_line.eps}
%}
%\label{fig3_55_75}
%\end{figure*}
\begin{figure}[h!]
\centerline{
\includegraphics[scale=0.8]{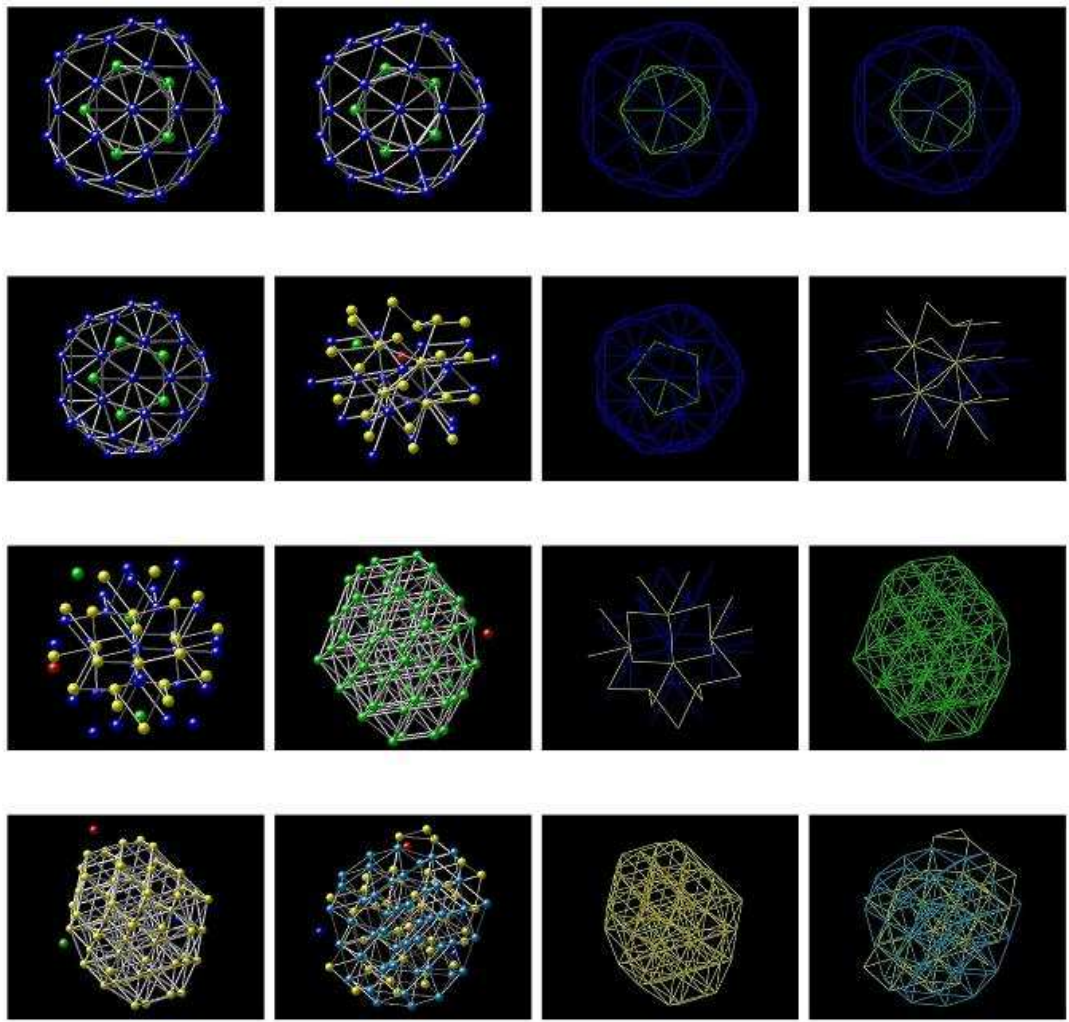}
}
\caption{N=39, 40, 42, 48, 55, 75, 76, 97 (got by Algorithm 2).}
\label{fig3}
\end{figure}

%\newpage
%\begin{figure*}[h!]
%\centerline{
%\includegraphics[scale=0.26]{Fig04_17_39atoms_others.eps}
%\includegraphics[scale=0.26]{Fig04_18_40atoms_others.eps}
%\includegraphics[scale=0.26]{Fig04_19_39atoms_others_line.eps}
%\includegraphics[scale=0.26]{Fig04_20_40atoms_others_line.eps}
%}
%\label{fig4_39_40}
%\end{figure*}
%\begin{figure*}[h!]
%\centerline{
%\includegraphics[scale=0.26]{Fig04_21_42atoms_others.eps}
%\includegraphics[scale=0.26]{Fig04_22_48atoms_others.eps}
%\includegraphics[scale=0.26]{Fig04_23_42atoms_others_line.eps}
%\includegraphics[scale=0.26]{Fig04_24_48atoms_others_line.eps}
%}
%\label{fig4_42_48}
%\end{figure*}
%\begin{figure*}[h!]
%\centerline{
%\includegraphics[scale=0.26]{Fig04_25_55atoms_others.eps}
%\includegraphics[scale=0.26]{Fig04_26_75atoms_others.eps}
%\includegraphics[scale=0.26]{Fig04_27_55atoms_others_line.eps}
%\includegraphics[scale=0.26]{Fig04_28_75atoms_others_line.eps}
%}
%\label{fig4_55_75}
%\end{figure*}
\begin{figure}[h!]
\centerline{
\includegraphics[scale=0.8]{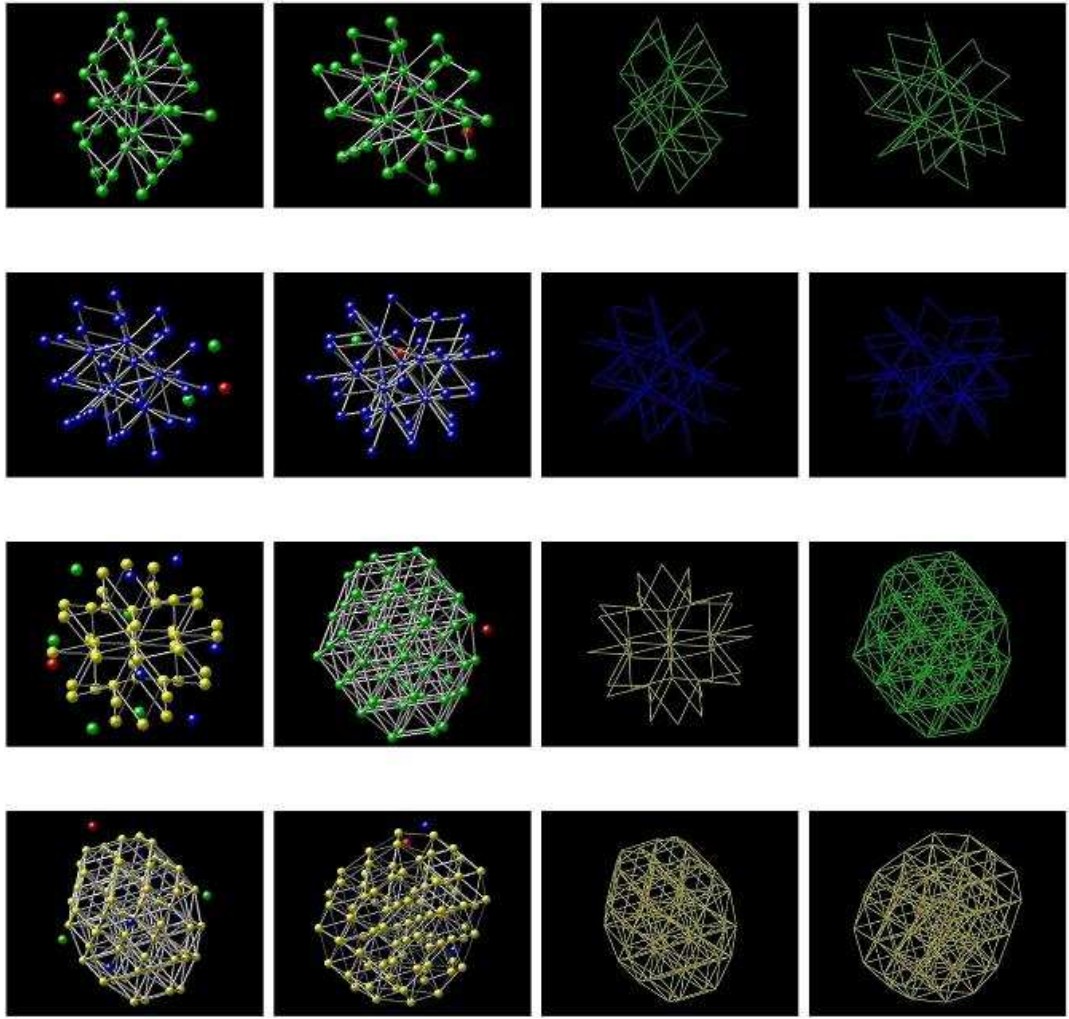}
}
\caption{N=39, 40, 42, 48, 55, 75, 76, 97 (best structures known).}
\label{fig4}
\end{figure}

In implementing Algorithm 1, we use T = 0.9*T as the temperature annealing schedule and the initial temperature is taken large enough according to the rule in \cite{kirkpatrick1983}, where we take $\text{T}_0$=300 K. We restrict the number of iterations for the outer procedure by 100 and number of iterations for the inner procedure by 1,000. The DG method part is terminated when the distance between the approximation to the subdifferential and origin is less than a given tolerance $\epsilon > 0$ ($\epsilon =10^{-4}$). The proposed hybrid method fails to solve (4) when number of atoms $N\geq 20$ (however, the AD Chains of Models 1-2 have 60/58 atoms). In order to reduce the number of local minima we suggest to approximate the function
\begin{equation} \label{tau_6-3}
\varphi(\tau) = 4(\frac{1}{\tau^6} -\frac{1}{\tau^3} )
\end{equation}
(which is neither convex nor strictly the difference of two convex functions) by the following function \cite{zhang2004}:
\begin{equation} \label{tau_max_min}
g(\tau)=\max (g_1(\tau),\min (g_2(\tau), g_3(\tau)))
\end{equation}
where $g_1(\tau)$ is the piecewise linear approximation of the function $\varphi(\tau)$ in segment $(0,r_0]$, $g_2(\tau)$ is the piecewise linear approximation of this function over segment $[r_0,r_1]$, and finally $g_3(\tau)$ is the piecewise linear approximation over $[r_1,b]$ and $b$ is an enough large number. Here
\begin{equation} \label{r_0-1}
r_0=\sqrt[3]{2},~~r_1=1/\sqrt[3]{2/7}.
\end{equation}
Such an approximation of the function $\varphi(\tau)$ allows us to remove many local minima of the Lennard-Jones potential function and to get a good approximation to the global minimum of the objective function $f$ in problem (4). In numerical experiments we take $b=16$ and divide the segment $[0.001,r_0]$ into 100 segments, the segment $[r_0,r_1]$ into 100 segments and the $[r_1,16]$ into 50 segments which allows one to get good approximations for the function $\varphi(\tau)$. The replacement of the function $\varphi(\tau)$ by the function $g(\tau)$ makes the objective function nonsmooth. On the other side such a replacement significantly reduce the number of local minima. Since the discrete gradient method is a method of nonsmooth optimization the proposed hybrid method can be applied for solving this transformed problem. When solving the L-J problem (4), first we use the DG method with build-up technique to relax to an initial solution. Then we apply the hybrid method, with the above approximation for the objective function, to get another initial solution. Starting from this initial solution we again apply the derivative-free DG method and at last get the global solution. Results of numerical experiments (Table 2) show that our techniques can effectively solve the L-J problem (4) when number of atoms is not greater than 310. For Model 1, seeing Figure 1 we may know that vdw interactions such as between 1D.CB–-6A.O, 3D.CB-–3A.CB, 6D.O-–1A.CB, etc. should be maintained. Solving the optimization problem (4) can get A Chain and D Chain, where D Chain should have good vdw interactions with A Chain. Similarly for Model 2, vdw interactions should be maintained between 3D.CB-3A.CB, etc. (Figure 2). AD Chains in all have 60/58 atoms. Thus, we may use the above improved hybrid SADG algorithm to easily get the optimal coordinates of AD Chains of Models 1-2, where D Chain has good vdw interactions with A Chain now (Figures 5-6). Other chains (i.e. $\beta$-strands) of Models 1-2 are got from AD Chains by the parallelization of AD Chains. The initial structures of Models 1-2 are shown in Figures 7-8.

%\noindent {\bf Table 2} Our numerical results for the L-J Potential Problem:
\begin{table}[h!]
\begin{center}
\caption{Our numerical results for the L-J Potential Problem}
\label{table2}\small{
\begin{tabular}{ccc}\hline
Number of atoms   &Best value obtained  &Best value known$^*$\\ \hline
19        &-72.659782        &-72.659782\\ \hline
20        &-77.177043        &-77.177043\\ \hline
21        &-81.684571        &-81.684571\\ \hline
22        &-86.573675        &-86.809782\\ \hline
23        &-92.844461        &-92.844472\\ \hline
24        &-97.348815        &-97.348815\\ \hline
25        &-102.372663       &-102.372663\\ \hline
27        &-112.825517       &-112.873584\\ \hline
30        &-128.096960       &-128.286571\\ \hline
34        &-150.044528       &-150.044528\\ \hline
44        &-207.631655       &-207.688728\\ \hline
49        &-239.091863       &-239.091864\\ \hline
56        &-283.324945       &-283.643105\\ \hline
65        &-334.014007       &-334.971532\\ \hline
67        &-347.053308       &-347.252007\\ \hline
84        &-452.267210       &-452.6573\\ \hline
93        &-510.653123       &-510.8779\\ \hline
148       &-881.072948       &-881.072971\\ \hline
170       &-1,024.791771     &-1,024.791797\\ \hline
172       &-1,039.154878     &-1,039.154907\\ \hline
268       &-1,706.182547     &-1,706.182605\\ \hline
288       &-1,850.010789     &-1,850.010842\\ \hline
293       &-1,888.427022     &-1,888.427400\\ \hline
298       &-1,927.638727     &-1,927.638785\\ \hline
300       &-1,942.106181     &-1,942.106775\\ \hline
301       &-1,949.340973     &-1,949.341015\\ \hline
304       &-1,971.044089     &-1,971.044144\\ \hline
308       &-1,999.983235     &-1,999.983300\\ \hline
\end{tabular}}
$^*${\small http://physchem.ox.ac.uk/$\sim$doye/jon/structures/LJ.html}
\end{center}
\end{table}

\begin{figure}[h!]
\centerline{
\includegraphics[scale=0.45]{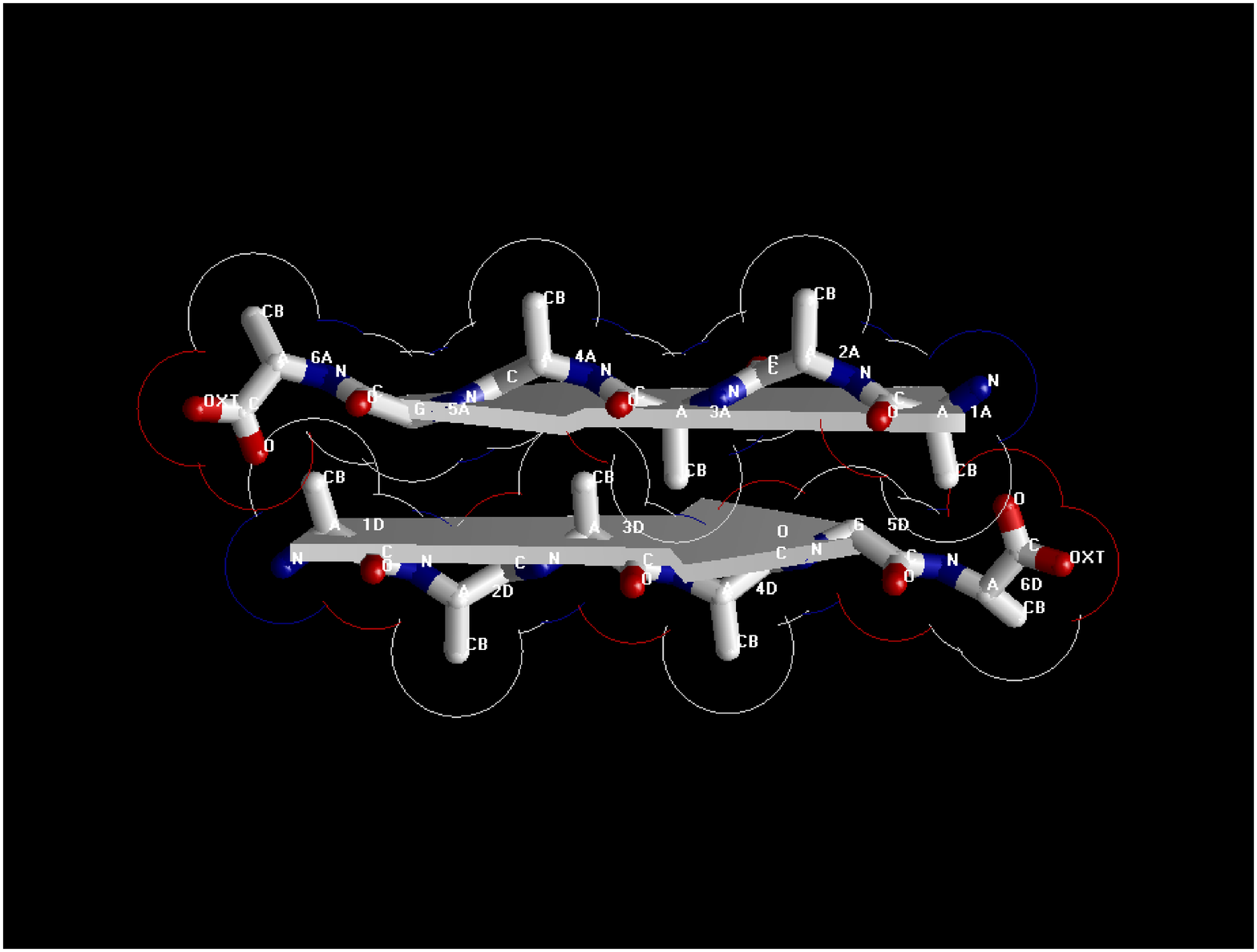}
}
\caption{Model 1 - good vdw interactions of AD chains of the AAAAGA model.}
\label{fig7}
\end{figure}
\begin{figure}[h!]
\centerline{
\includegraphics[scale=0.45]{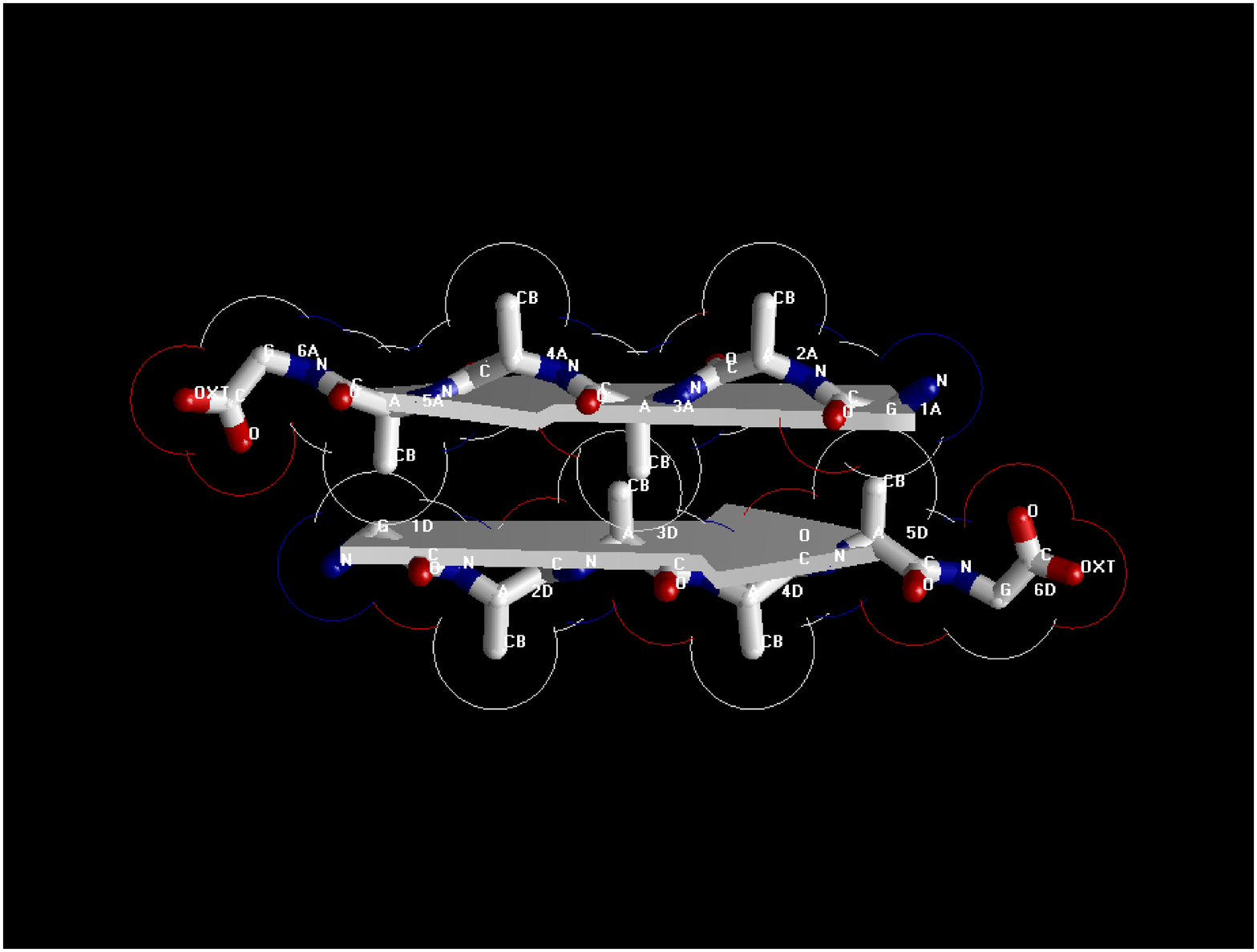}
}
\caption{Model 2 - good vdw interactions of AD chains of the GAAAAG model.}
\label{fig8}
\end{figure}

\begin{figure}[h!]
\centerline{
\includegraphics[scale=0.45]{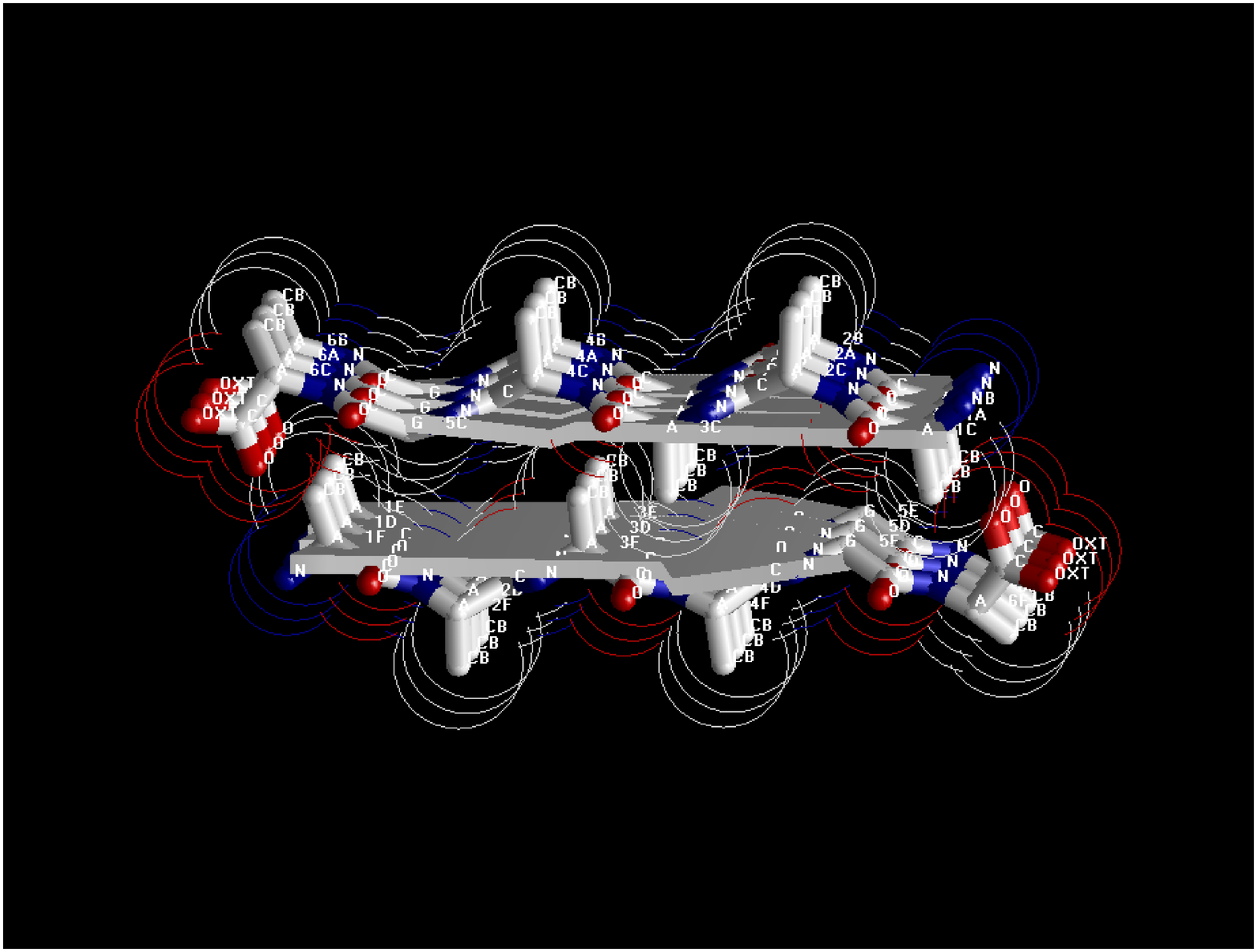}
}
\caption{Model 1 - initial structure of prion AAAAGA amyloid fibril.}
\label{fig9}
\end{figure}
\begin{figure}[h!]
\centerline{
\includegraphics[scale=0.45]{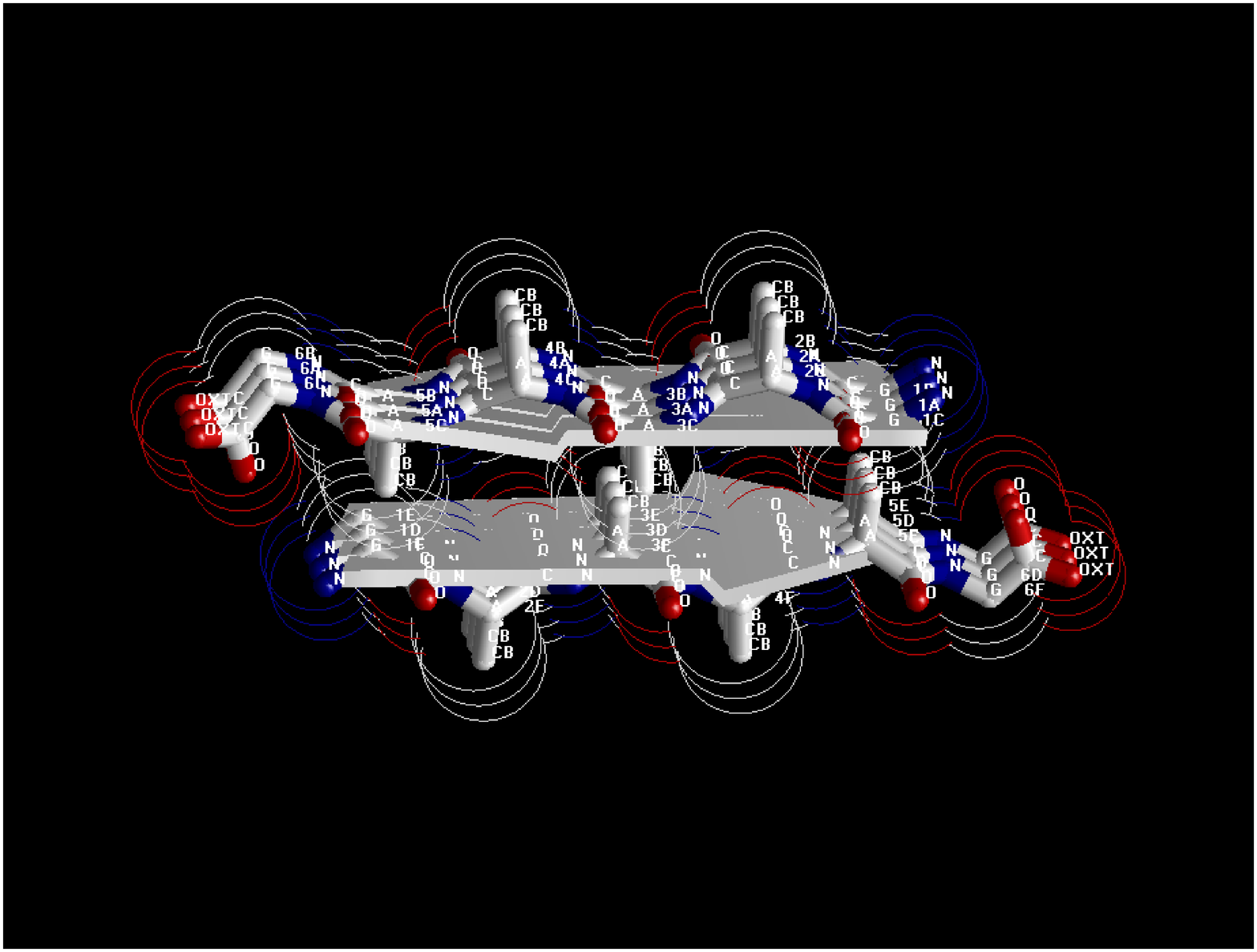}
}
\caption{Model 2 - initial structure of prion GAAAAG amyloid fibril.} \vskip 1cm
\label{fig10}
\end{figure}

\section{Model Solving/Optimization}
The L-J potential (2) energy of atoms' vdw interactions is just a part of the total potential energy of a protein \cite{case2008,locatelli2008}:
\begin{eqnarray}\label{potential}
E_{\rm total} = &\sum_{\rm bonds} K_r (r - r_{eq})^2 \nonumber\\
              &+\sum_{\rm angles} K_\theta (\theta - \theta_{eq})^2 \nonumber\\
              &+\sum_{\rm dihedrals} {V_n \over 2} [1 + {\rm cos}(n\phi - \gamma)] \nonumber\\
              &+\sum_{i<j}^{\rm vdw} \left [ {A_{ij} \over R_{ij}^{12}} - 
                                   {B_{ij} \over R_{ij}^6} 
                          \right ] \nonumber\\
              &+\sum_{i<j}^{\rm electrostatic} \left [ 
                                   {q_iq_j \over \epsilon R_{ij}} 
                          \right ] \nonumber\\
              &+\sum_{\rm H-bonds} 
                           \left [ {C_{ij} \over R_{ij}^{12}} - 
                                   {D_{ij} \over R_{ij}^{10}} 
                           \right ]. 
\end{eqnarray}
The initial structures of Models 1-2 illuminated in Figures 7-8 are not the optimal structures with the lowest total potential energies. The initial structures also have no hydrogen atoms (so no hydrogen bonds existed) and water molecules added. For each Chain, the C-terminal and N-terminal atoms also have problems. Clearly there are a lot of close/bad contacts between $\beta$-strand atoms as illuminated in Figures 7-8. Thus,  we still use the hybrid techniques of SD, CG, SA optimization methods within AMBER \cite{case2008,zhang2011a} to optimize the above Models 1-2 in order to get the most stable structures. Each of the most stable structures will have its lowest total potential energy, i.e.
\begin{equation}
\min E_{\rm total}.
\end{equation}

We used the ff03 force field of AMBER 10, in a neutral pH environment. The amyloid fibrils were surrounded with a 8 angstrom layer of TIP3PBOX water molecules using the XLEaP module of AMBER 10. 1,360, 1,372 waters and 180, 168 hydrogen atoms were added separately for Models 1-2 by the XLEaP module. The solvated amyloid fibrils were minimized by the SD method and then the CG method were performed (OPT1). Model 1 were optimized by 95,016 steps of SD and 27,751 steps of CG; Model 2 by 95,016 steps of SD and 24,418 steps of CG. Then the solvated amyloid fibrils were quickly heated from 0 K to 300 K linearly during 20 ps. The systems were kept at 300 K for 80 ps. The systems then were slowly cooled from 300 K to 100 K linearly for 400 ps. At 100 K, the systems were kept for 100 ps, and then for 4,400 ps until the systems reach sufficient equilibration at 100 K (the RMSD, PRESS, and VOLUME (DENSITY) were sufficiently stable though their variations are very large). The \underline{SA}NDER (Simulated Annealing with NMR-Derived Energy Restraints) algorithm with nonbonded cutoffs of 12 angstroms were used during the heating, cooling and the 100 ps at 100 K. Step size is 2 fs for the whole SA runs. During the SA, the Metropolis criterion was used. After the SA, the models were refined by SD and CG methods again (OPT2), Model 1 was refined by 20,000 steps of SD and 597 steps of CG, and 20,000 steps of SD and 1,921 steps of CG for Model 2. All the above works were performed on the Tango facilities of the Victorian Partnership for Advanced Computing (http://www.vpac.org) of Australia. 

\section{Results and Discussion}
Figures 9-10 show the potential energy development for the two Models (where the OPT1-SA-OPT2 of AMBER 10 were used to generate the potential energy and the Figures were drawn with XMGRACE of Grace 5.1.21). We can see that the potential energy goes down during the SD and CG optimization phase OPT1, suddenly drops down and quickly goes up and then slowly goes down and levels off during the SA phase, and at last quickly goes down and then levels off during a short phase of OPT2. At the beginning of SA, the energy quickly drops off is due to the temperatures of the systems being suddenly changed from 100 K \cite{wiltzius2009} to 0 K. This is a case of so called ``quenching". Some energy values are listed in Table 3. In Table 3, the first column of energies (OPT1 1st-step) are the ones of the initial structures of Models 1-2. The distance between $\beta$-strands is too short for the vdw contacts so that Amber 10 cannot show the large L-J potential values (in Table 3 Column 1). This also implies the initial structures (Figures 7-8) are far from their optimal structures. OPT1 removes these bad vdw and hydrogen bond contacts and makes the structures become much better with lower potential energies. However, OPT1 is a local search optimization method which cannot thoroughly optimize the models into their most stable structures.

\begin{figure}[h!]
\centerline{
\includegraphics[scale=0.5]{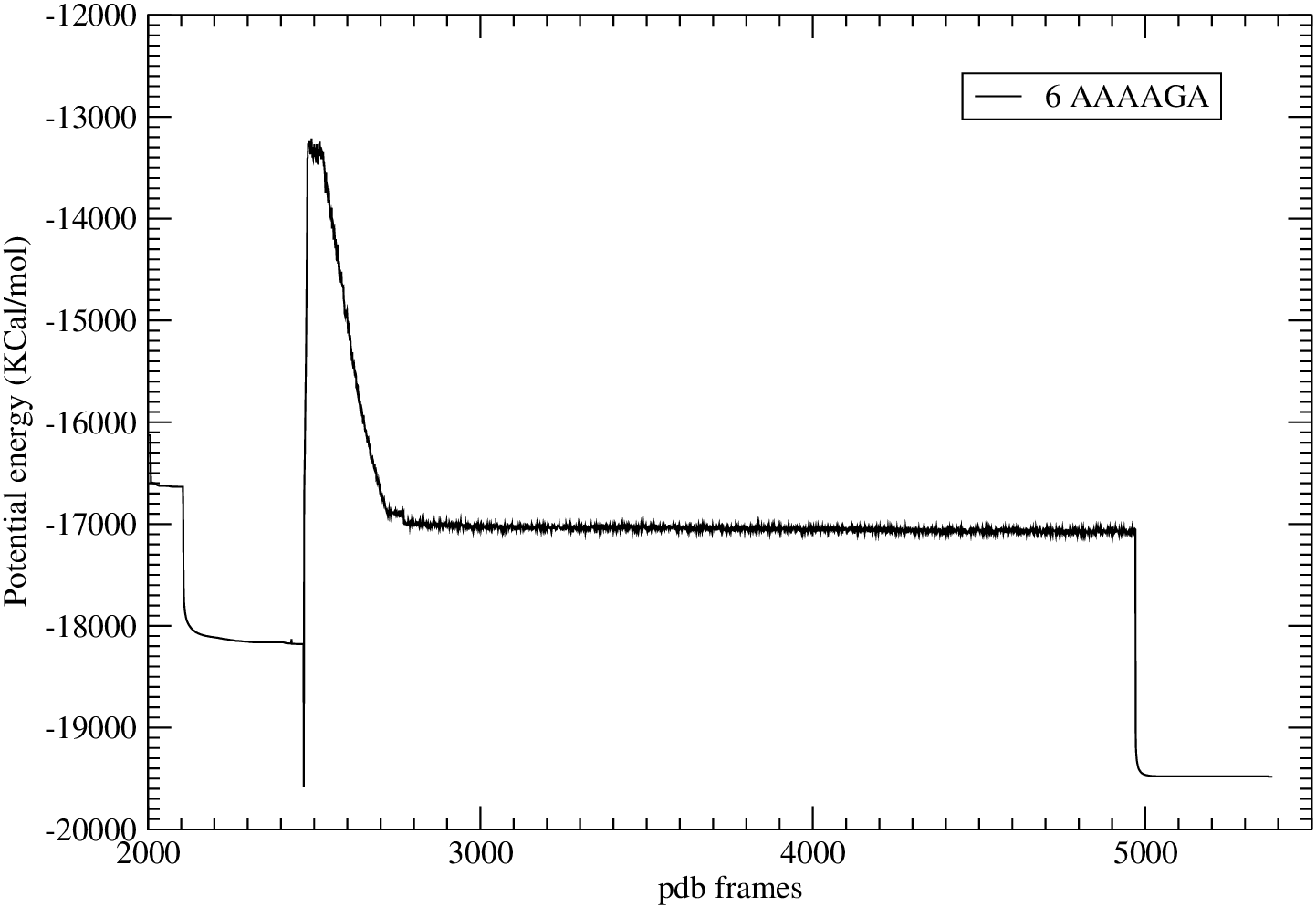}
}
\caption{Potential energy of Model 1.} \vskip 1.2cm
\label{fig11}
\end{figure}

\begin{figure}[h!]
\centerline{
\includegraphics[scale=0.5]{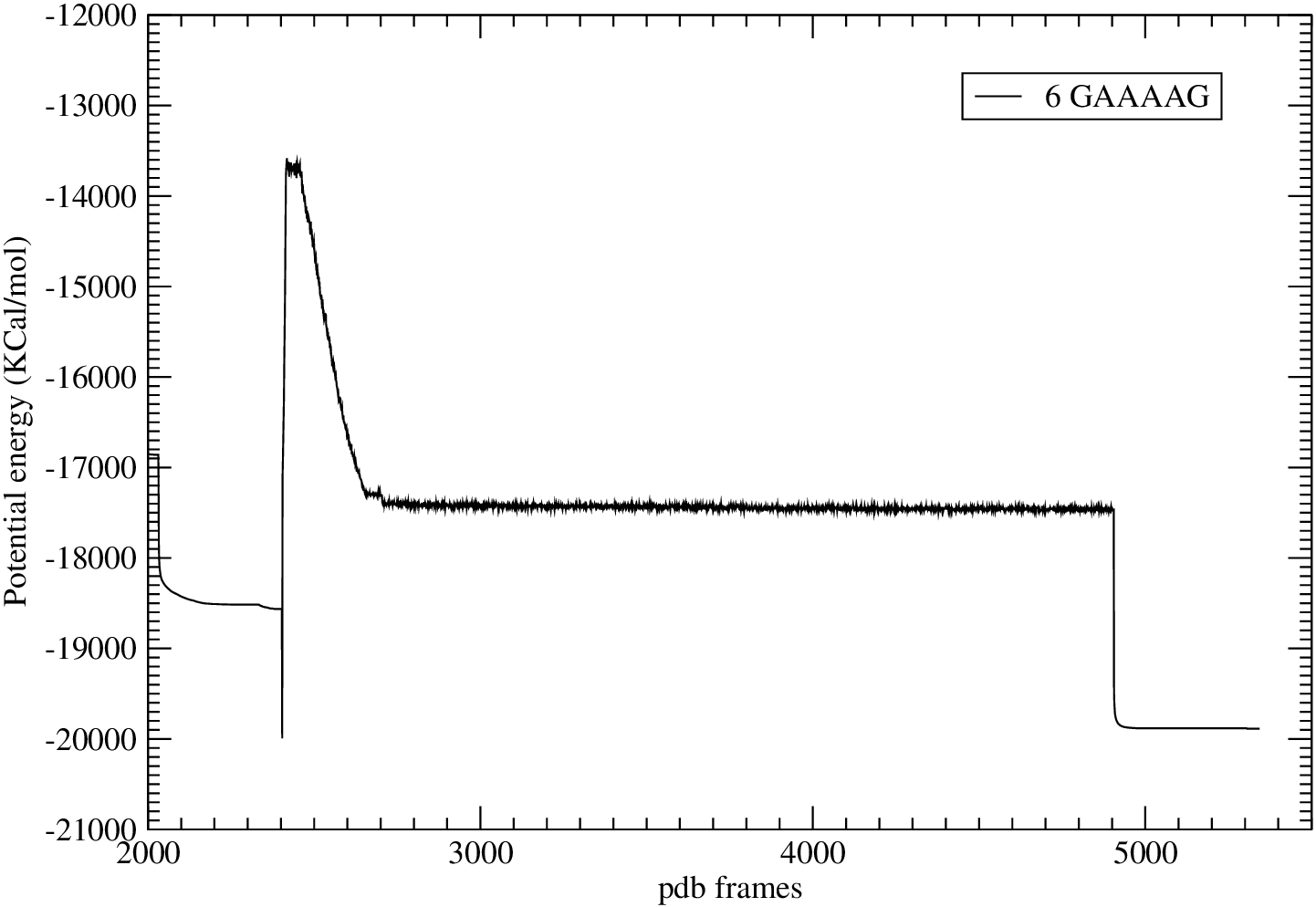}
}
\caption{Potential energy of Model 2.}
\label{fig12}
\end{figure}

%\noindent {\bf Table 3} Potential energy values:
\begin{table}[h!]
\begin{center}                 
\caption{Potential energy values}
\label{table3} {\tiny
\begin{tabular}{lccccccc}
\\ \hline      
Model  &OPT1 1st-step &OPT1 50th-step &OPT1 last-step     &SA 1st-step &SA Last 10,000-steps  &OPT2 1st-step &OPT2 last-step\\ 
       &              &               &                   &            &(average value)      &              &              \\ \hline   
1      &5.0294E+12    &117,540        &-18,179            &-19,583.8427&-17,074.54472        &-17,053       &-19,482\\
vdw    &******      &36,804.4917    &                   &            &                     &              &\\ \hline
2      &1.8427E+16    &419,080        &-18,564            &-19,992.2174&-17,461.46262        &-17,474       &-19,887\\
vdw  &******        &232,373.1312   &                   &            &                     &              &\\ \hline 
\end{tabular}}
\end{center}
\end{table}

In Figures 9-10 we see that models are trapped into their local optimal structures. SA is a global search optimization method that can make OPT1 jump out of the local trap, even accepting very bad cases with low probability according to the Metropolis criterion. Thus, in Table 3 we see that SA rapidly quenches the molecular structures, allowing escape from the local traps; SA finally results in the loss of 2,509.29798 KCal/mol, 2,530.75478 KCal/mol for the two systems, respectively.\\

After SA, OPT2 can safely bring the molecular structures of the models to the most stable states. OPT2 makes the molecules in Models 1–2 lose 2,429 KCal/mol, 2,413 KCal/mol of potential energy, respectively. OPT2 results in a loss of energy from Models 1–2 of nearly the same magnitude as that of SA (i.e. the decrease in energy in OPT2 is significant compared to the change of energy in SA between the 1st step and the average of the last 10,000 steps). OPT1 could not make further optimization, but OPT2 could make further optimization after SA; this demonstrates the effectiveness of SA (shown in Table 3 by comparing the values of OPT2 last-step with OPT1 last-step).\\

The final optimal molecular structures of Models 1-2 after OPT2 are shown in Figures 11-12, where the snapshots after OPT2 were drawn by the free package Molecular Visualisation \& Modeling (MVM) (http://www.zmmsoft.com/). The RMSDs (root mean square deviations) from the initial structures shown in Figures 7-8 (where the initial structures were drawn by MVM and were generated by the improved SADG Algorithm 1) are 2.71, 2.95 angstroms, respectively, for Models 1-2. The hydrogen bonds between the two closet adjacent $\beta$-strands and the vdw contacts between the two inner closest adjacent alanines can be seen in Figures 11-12. In both models, there is about 5 angstroms between the two closet adjacent $\beta$-sheets, maintained by hydrophobic bonds, and about 4.5 angstroms between the two closet adjacent $\beta$-strands, which are linked by hydrogen bonds such as Ala2.H-Ala7.O, Ala6.H-Gly11.O for Model 1 and Ala2.O-Ala8.H, Ala21.H-Ala32.O for Model 2. There is a hydrophobic core in each of the models. These amyloid fibrils are rich in $\beta$-sheet structure and contain a cross-$\beta$ core form of infectious prions, which causes prion diseases.

\begin{figure}[h!]
\centerline{
\includegraphics[scale=0.45]{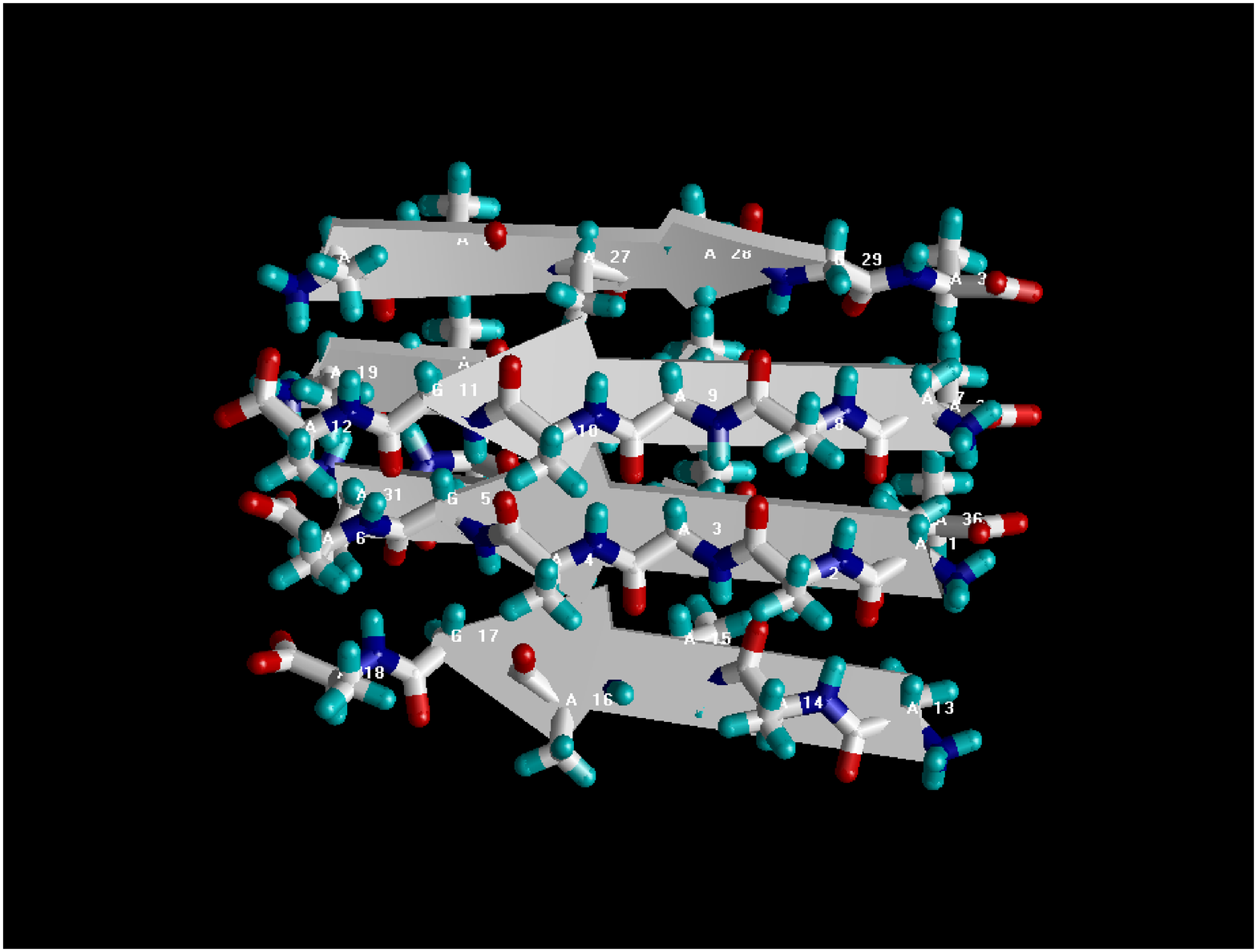}
}
\caption{Model 1 - optimal structure of prion AAAAGA amyloid fibril.}
\label{fig13}
\end{figure}

\begin{figure}[h!]
\centerline{
\includegraphics[scale=0.45]{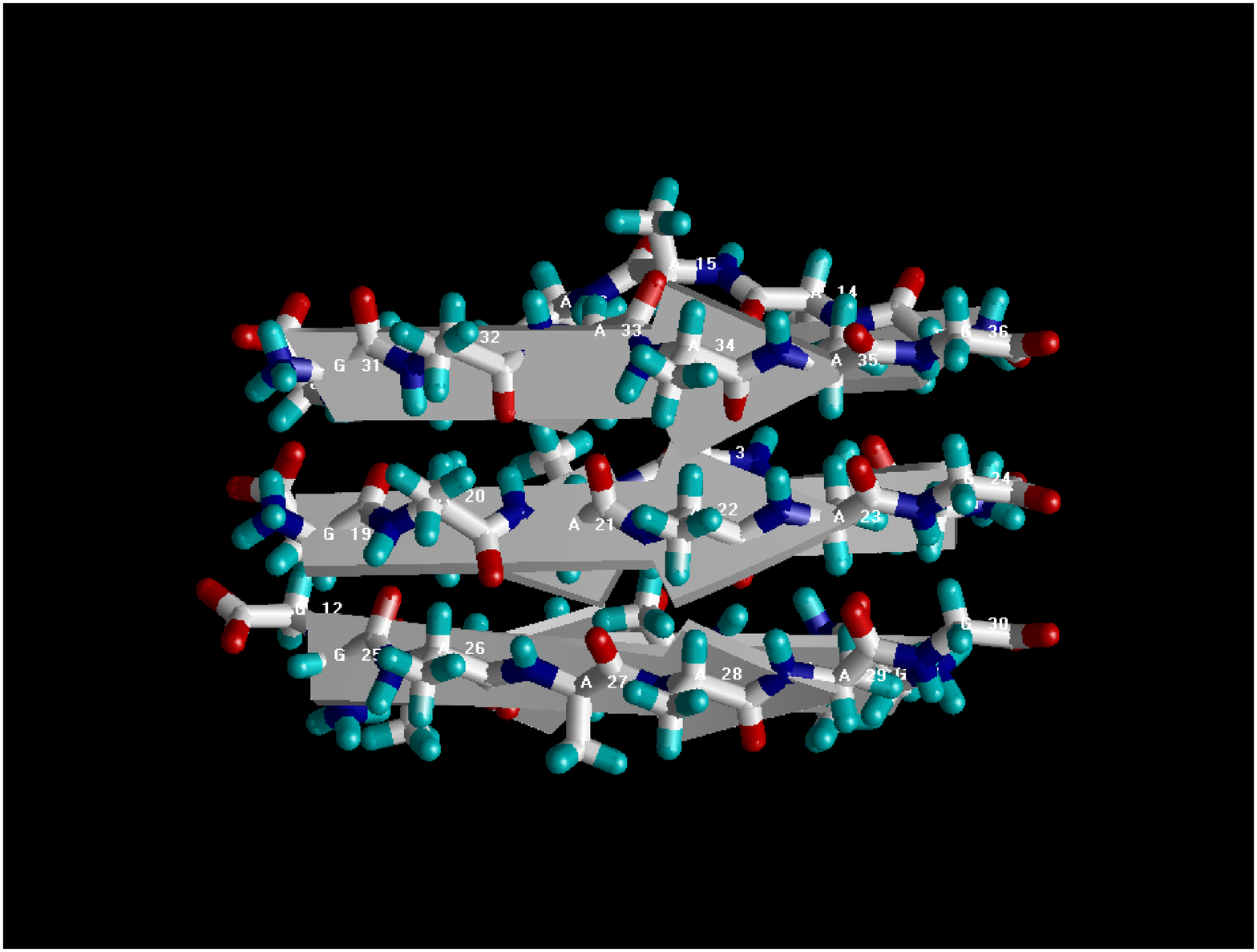}
}
\caption{Model 2 - optimal structure of prion AAAAGA amyloid fibril.}
\label{fig14}
\end{figure}

Numerical results of this paper showed that a six chains AGAAAA model could not successfully pass SA. However, two prion AGAAAAGA palindrome amyloid fibril models - a six chains AAAAGA model (Model 1) and a six chains GAAAAG model (Model 2) - were successfully passing OPT1-SA-OPT2 and got at last.

\section{Conclusion}
In recent years large-scale global optimization (GO) problems have drawn considerable attention. These problems have many applications, in particular in biochemistry and data mining. Numerical methods for GO are often very time consuming and could not be applied for high-dimensional non-convex and/or non-smooth optimization problems. The study of new algorithms which allow one to solve large-scale GO problem is very important. One technique is to use hybrid of global and local/global search algorithms. This paper presents two hybrid methods for solving the large-scale L-J potential GO problem. The methods do not guarantee the calculation of a global solution; however results of numerical experiments show that they, as a rule, calculate a solution which is global one or close to it. The improved hybrid SADG method can be successfully applied to the construction work of optimal atomic-resolution structures of prion AGAAAAGA amyloid fibrils. As the three models constructed for amyloid fibrils in \cite{zhang2011a}, the two amyloid fibril models gained in this paper may be useful in furthering the goals of medicinal chemistry.

{\small
\section*{Acknowledgments}
The authors are grateful to Professor Adil M. Bagirov (University of Ballarat, Australia) for making available FORTRAN code of his Discrete Gradient Method and for discussions of some ideas on solving the Lennard-Jones Potential Optimization Problem. This research was supported by a Victorian Life Sciences Computation Initiative grant No. VR0063; the authors would like to thank the staff of Universities of Ballarat, Curtin and Melbourne for their supports and helps to this project. This paper is dedicated to Professor Kok Lay Teo (Curtin University of Technology, Australia) and Professor Jie Sun (National University of Singapore, Singapore) on their 65th birthday and to Professor Changyu Wang (Qufu Normal University, P. R. China) on his 75th birthday, to Professor Masao Fukushima (Kyoto University, Japan) on his 60th birthday; thanks go to them for their supports in the last 20 years. Thanks also go to Dr. Honglei Xu (Curtin University of Technology, Australia) et al. for getting the motive to write this paper for the International Conference on Optimization and Control 2010, July 2010, Perth/Chongqing/Guiyang (http://sci.gzu.edu.cn/icoco/ICOCO-Proceedings.pdf,  pp. 91--112). Last, but not least, the authors appreciate the anonymous referees for their numerous insightful comments to improve this paper.
}

\end{document}